\documentclass{ws-mpla}

\def    \bea           {\begin{eqnarray}}
\def    \eea           {\end{eqnarray}}

\def\lsim{\raise0.3ex\hbox{$\;<$\kern-0.75em\raise-1.1ex\hbox{$\sim\;$}}}
\def\gsim{\raise0.3ex\hbox{$\;>$\kern-0.75em\raise-1.1ex\hbox{$\sim\;$}}}

\begin{document}

\markboth{Carlos Mu\~noz}
{A kind of prediction from string phenomenology: extra matter at low energy}

\catchline{}{}{}{}{}

\title{A KIND OF PREDICTION FROM
STRING PHENOMENOLOGY:\\
EXTRA MATTER AT LOW ENERGY 
}

\author{\footnotesize CARLOS MU\~NOZ
}

\address{Departamento de F\'\i sica Te\'orica C-XI and Instituto de F\'\i sica
  Te\'orica C-XVI,\\ 
Universidad Aut\'onoma de Madrid,
Cantoblanco, E-28049 Madrid, Spain\\
carlos.munnoz@uam.es}



\maketitle


\begin{abstract}
We review the possibility that the Supersymmetric Standard Model arises
from orbifold constructions of the $E_8\times E_8$ Heterotic Superstring,
and the phenomenological properties that such a model should have.
In particular, 
trying to solve the discrepancy between the
unification scale predicted by the Heterotic Superstring
($\approx g_{GUT}\times 5.27\cdot 10^{17}$ GeV) 
and the value deduced from LEP experiments 
($\approx 2\cdot 10^{16}$ GeV), we will predict 
the presence at low energies of 
three families of Higgses
and vector-like colour triplets. 
Our approach relies on the Fayet-Iliopoulos breaking, and this is also
a crucial ingredient, together with having three Higgs families, 
to obtain in these models an interesting pattern of 
fermion
masses 
and mixing angles at the renormalizable lebel. Namely, after the gauge breaking some physical
particles
appear combined with other states, and the Yukawa couplings are
modified 
in a well controlled way.
On the other hand, dangerous
flavour-changing neutral currents may appear when
fermions of a given charge receive their mass through couplings
with several Higgs doublets.
We will address this potential problem, finding that
viable scenarios can be obtained for a reasonable light Higgs spectrum.

\end{abstract}

\ccode{PACS Nos.: 11.25.Wx, 11.25.Mj, 12.60.Jv, 12.60.Fr}

\section{Introduction}

In SuperString Theory the elementary particles are not point-like objects but
extended, string-like objects. It is surprising that this 
apparently 
small change allows us to answer fundamental questions that in the 
context
of the quantum field theory of point-like particles cannot even be posed.
For example: 
Why is the Standard Model gauge group
$SU(3)\times SU(2)_L\times U(1)_Y$?
Why are there three families of particles?
Why is the mass of the electron $m_e=0.5$ MeV? 
Why is the fine structure constant $\alpha=1/137$?
In addition, only SuperString Theory 
has the potential to unify all gauge interactions 
with gravity 
in a consistent 
way.
In this sense, it is a crucial step in the construction of the
fundamental theory of particle physics
to find a consistent SuperString model in four dimensions
accommodating the observed Standard Model (SM), 
i.e. we need to find the SuperString Standard Model (SSSM).
Actually, this is the main task of what we call String Phenomenology.

In the late eighties,
the compactification of the $E_8\times E_8$
Heterotic String 
on six-dimensional orbifolds 
proved to be an
interesting method to carry out this 
task\footnote{Other interesting attempts 
at model building used Calabi--Yau spaces
and fermionic constructions.
} 
(for a brief historical account see the Introduction in 
Ref.~\refcite{prediction}
and references therein).
For example, it was shown that the use of two 
Wilson lines 
on the torus defining the
symmetric $Z_3$ orbifold can give rise to four-dimensional
supersymmetric models with gauge group
$SU(3)\times SU(2)\times U(1)^5\times G_{hidden}$ 
and, automatically, three generations of chiral particles\cite{Kimmm}. 
In addition,
it was also shown 
that the 
Fayet--Iliopoulos (FI) D-term, 
which appears because of the presence of 
an anomalous $U(1)$, can give
rise to the breaking of the extra $U(1)$'s.
In this way it was possible to construct\cite{Casas2,Font,Katehou} 
supersymmetric 
models 
with gauge group 
$SU(3)\times SU(2)\times U(1)_Y$,
three generations of particles in the observable sector,
and absence of dangerous 
baryon- and 
lepton-number-violating operators\footnote{Recently,
other interesting models in the context of the $Z_3$ orbifold\cite{Giedt2},
as well as in the context of the $Z_6$ orbifold\cite{Raby,Lebedev,Lebedev2},
and
$Z_{12}$ orbifold\cite{Kim,hum},
have been analysed.}.
 

Unfortunately, we cannot claim that one of these $Z_3$ orbifold 
models is the 
SSSM, since several problems are always present. 
For example, 
the initially large number of extra particles,
which are generically present in these constructions, is 
highly reduced through the
FI mechanism,
since many of them get a high mass 
($\approx 10^{16-17}$ GeV). However, in general,
some extra $SU(3)$ triplets, $SU(2)$ doublets and 
$SU(3)\times SU(2)$ singlets still remain at low energy.
On the other hand,
given the predicted value for the unification scale in
the Heterotic String\cite{Kaplu}, 
$M_{GUT}\approx g_{GUT}\times 5.27\cdot 10^{17}$ GeV,
the values of the gauge couplings deduced from
LEP experiments cannot\cite{rges} be obtained\footnote{Recall 
that this is only possible 
in the context of the Minimal Supersymmetric Standard Model (MSSM)
for $M_{GUT}\approx 2\times 10^{16}$ GeV.}.
It was also not possible to obtain in these models the necessary 
Yukawa couplings
reproducing the observed fermion masses\cite{Katehou,Font,Casas3}. 

At this point, it is fair to say
that almost 20 years have gone by since String Phenomenology started,
and the SSSM has not been found yet\footnote{And this sentence can also be
applied to any of the interesting models constructed in more
recent years using D-brane technology\cite{dbranes}.
Actually, the probability of obtaining an MSSM like gauge group with
three generation 
in the context of intersecting D-branes in an orientifold background
seems to be extremelly small\cite{proba,proba2}, of about $10^{-9}$.}
As acquittal on the charge we should remark that there are thousands of models 
(vacua) that can be built.
Some of them have
the gauge group of the SM or GUT groups,
three families of particles, and other interesting properties, but
many others have 
a number of families different from three, 
no appropriate gauge groups, no appropriate matter, etc.
A perfect way of solving this problem would be to use
a dynamical mechanism to select the correct model (vacuum).
Such a mechanism should be able to determine a 
point in the parameter space of the Heterotic String
determining the correct
compactification
producing the gauge group
$SU(3)_c\times SU(2)_L\times U(1)_Y$, three families of the known particles, the
correct Yukawa couplings, etc.
The problem is that such a mechanism has not been discovered yet.

So, for the moment, the best we can do is keep trying,
i.e. to use the experimental results available (such as 
the SM gauge group, three families,
fermion masses, mixing angles, etc.), 
to discard models.
Although the model space is in principle huge, a detailed analysis
can reduce this to a reasonable size.
For example, within the 
$Z_3$ orbifold with two Wilson lines,
one can construct in principle a number 
of order 50000 of three-generation 
models
with the
$SU(3)\times SU(2)\times U(1)^5$ gauge group 
associated to the first $E_8$ of the Heterotic String.
However, a study implied that most of them are equivalent\cite{Mondragon}, 
and in fact, at the end of the day, only 192 different models were
found\cite{Giedt,Mondragon}.
This reduction is remarkable, but we should keep in mind that
the analysis of each one of these models is really complicated.

Nevertheless, a certain degree of optimism is important when working
in String Phenomenology, and one can argue that if the
SM arises from SuperString Theory there must exist one model
with the right properties.
In the present review we will adopt this viewpoint, and will assume
that the SM arises from orbifolds constructions.
Instead of the painful work of searching for the correct orbifold model, 
we will try to deduce the phenomenological properties 
that such a model must have
in order to solve the crucial problems
mentioned above, with the hope that this analysis will allow us
to make predictions that can be tested at the LHC.


In fact, all those problems,
extra matter, gauge coupling unification, and correct Yukawa couplings,
are closely related.
The first two because
the evolution of the gauge couplings from high to low energy
through the
renormalization group equations (RGEs),
depends on the existing matter\cite{mass,Giedt2}.
In Section~2 we will discuss a solution to the gauge coupling unification problem 
implying 
the prediction of three generations of supersymmetric Higgses
and vector-like colour triplets
at low energies\cite{prediction}. In this solution the FI scale plays
an important role.

Concerning the third problem, how to obtain
the observed structure of
fermion masses and mixing angles,
this is in our opinion
the most difficult task in String Phenomenology.
For example,
the right model must reproduce also the correct
mass hierarchy for quarks and leptons,
$\frac{m_t}{m_u}\sim 10^5$, $\frac{m_{\tau}}{m_e}\sim 10^3$, etc.,
and this is not a trivial task,
although
it is true that
one can find interesting results in the literature.
In particular, orbifold spaces have a beautiful mechanism to generate
a mass hierarchy at the renormalizable level.
Namely, Yukawa couplings between twisted matter can be explicitly computed
and they get suppression factors, 
which depend on the
distance between the fixed points to which the relevant fields are
attached\cite{Hamidi}$^-$\cite{Abel}. The couplings 
can be schematically written as
$\lambda\sim e^{-\sum_i c_{\lambda}^{i} T_i}$, with $Re\ T_i\sim R_i^2$, and 
the $T_i$ are the moduli fields associated to the size and shape of the
orbifold. The distances can be varied by giving different 
vacuum expectation values (VEVs) to these
moduli, implying that one can span in principle five orders of magnitude the
Yukawa couplings\cite{test}$^-$\cite{Abel}.
Unfortunately, this is not the end of the story, since Nature tells us
that
a weak coupling matrix exists with weird magnitudes for the entries, 
and that therefore we must arrange our up-and down-quark Yukawa
couplings in order to have specific off diagonal 
elements\footnote{
Needless to say, the recent experimental confirmation of neutrino
masses makes
the task even more involved. We have to explain also the weak coupling
matrix with the charged leptons.
Besides, in addition to the hierarchies shown above, we have to explain
others such as $\frac{m_e}{m_\nu }\gsim 10^6$.}.
In Section 3 we will see that to obtain this {\it at the renormalizable
level} is possible if three Higgs families and the FI 
breaking are
present\cite{Abel,teixeira2}.
Thus we have a common solution for the three problems mentioned above. 


On the other hand, it is well known that dangerous
flavour-changing neutral currents (FCNCs) may appear when
fermions of a given charge receive their mass through couplings
with several Higgs doublets\cite{Paschos,Paschos2}.
This situation might be present here since we have three generations
of supersymmetric Higgses.
In Section~4 we will address this potential problem, finding that
viable scenarios can be obtained\cite{teixeira2}.



\section{Predictions from gauge coupling unification
}


Since we are interested in the analysis of gauge couplings,
we need to first clarify which is the relevant scale for
the running between the supersymmetric scale $M_S$ and the unification
point.
Let us recall that in heterotic compactifications
some scalars singlets $C_i$ 
develop vacuum expectation values (VEVs) in order
to cancel the FI $D$-term, without breaking the
SM gauge group.
An estimate about their VEVs can be done
with the average 
result $\langle C_i \rangle\sim 10^{16-17}$ GeV (see e.g. Ref.~\refcite{Giedt2}).
After the breaking, many particles, say $\xi$, acquire a high mass because
of the generation of effective mass terms. These come for example
from operators of the type $C_i\xi\xi$.
In this way extra vector-like triplets and doublets and also singlets become 
very heavy. 
We will use the above value as our relevant scale,
the so-called
FI scale $M_{FI}\approx 10^{16-17}$ GeV.

As discussed in the Introduction, 
we are interested in the unification of the gauge couplings
at
$M_{GUT}\approx g_{GUT}\times 5.27\cdot 10^{17}$ GeV.
This is not a simple issue, and various approaches towards 
understanding it have been proposed in the literature\cite{Dienes}.
Some of these proposals consist of using string GUT models,
extra matter at
intermediate scales, heavy string threshold corrections, non-standard
hypercharge normalizations, etc.
In our case, we will try  
to obtain this value by using first the existence of extra matter at the scale
$M_S$. We will see that this is not sufficient and, as a consequence, 
the FI scale 
must be included. 
Let us concentrate for the moment on
$\alpha_3$ and  $\alpha_2$.
Recalling that three generations appear automatically 
for all the matter in $Z_3$ orbifold scenarios with two Wilson lines,
the most natural possibility
is to assume the presence of three light generations of supersymmetric
Higgses.
This implies that we have four extra Higgs doublets, $n_2=4$, 
with respect to the
case of the MSSM. 
Unfortunately, this goes wrong.
Whereas $\alpha_3^{-1}$ remains unchanged, since the number of extra
triplets is
$n_3=0$,
the line 
for $\alpha_2^{-1}$ is pushed 
down with respect to the case of the MSSM. 
As a consequence, the two
couplings cross at a very low scale
($\approx 10^{12}$ GeV).
We could try to improve this situation by assuming the presence of
extra triplets in addition to the four extra doublets. 
Then the line for $\alpha_3^{-1}$ is
also pushed down and therefore the crossing might be obtained
for larger scales. However, even for the
minimum number of extra triplets that can be naturally obtained
in our scenario, $3\times \{(3,1)+(\bar 3, 1)\}$, i.e. $n_3=6$,
the ``unification'' scale turns out to be too large ($\approx 10^{21}$
GeV).
One can check that
other possibilities including more extra doublets and/or triplets
do not work\cite{prediction}.
Thus, using extra matter at $M_S$ we are not able to obtain the 
Heterotic String unification scale, since
$\alpha_3$ never crosses $\alpha_2$ at
$M_{GUT}\approx g_{GUT}\times 5.27\cdot 10^{17}$ GeV.
Fortunately, this is not the end of the story. 
As we will show now,
the FI scale $M_{FI}$ is going to
play an important role in the analysis.

In order to determine whether or not the  
Heterotic String unification scale
can be obtained, we need to know the number of doublets
$n_2^{FI}$ and triplets $n_3^{FI}$ 
in our construction with masses of order the FI
scale $M_{FI}$.
It is possible to show
that within
the $Z_3$ orbifold with two Wilson lines, three-generation 
standard-like models must fulfil the following relation
for the extra matter:
$2+n_2+n_2^{FI}=n_3+n_3^{FI}+12$.
Then, it is now straightforward to check
that only models with 
$
n_2=4, n_3=6, 
$
%
and therefore 
$n_2^{FI}-n_3^{FI}=12$, may give rise to the Heterotic String
unification scale\cite{prediction} 
(other possibilities for $n_2$, $n_3$ 
do not even produce the crossing of $\alpha_3$ and $\alpha_2$). 
This is shown in Fig.~1 for an example with
$n_3^{FI}=0$, and assuming $M_S=500$ GeV. There we are
using $M_{FI}=2\cdot 10^{16}$ GeV as will be discussed below. 

\begin{figure}[t]
\centerline{\epsfxsize=2.5in\epsfbox{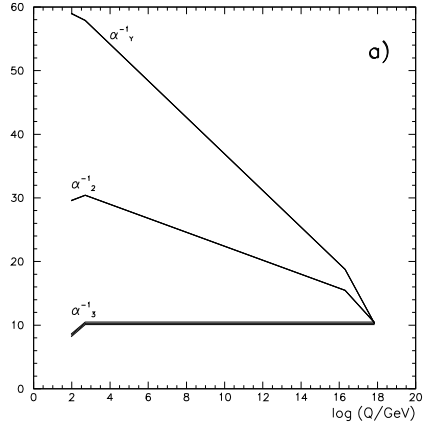}}   
\caption{
Unification of the gauge couplings 
at $M_{GUT}\approx g_{GUT}\times 5.27\cdot 10^{17}$ GeV 
with three light generations of supersymmetric Higgses
and
vector-like colour triplets.
In this example we show one of the four 
possible patterns of heavy matter in eq.~(\ref{extra5}),
in particular that with {\it a)} $n_3^{FI}=0$.
The line corresponding to $\alpha_1$
is just one of the many possible examples.
}
\end{figure}



Note that at low energy we then have (excluding singlets)
\bea
3\times \left\{(3,2)+2(\bar 3,1)+(1,2)\right\}
+ 3\times \left\{(3,1)+(\bar 3,1)+ 2(1,2)\right\}
\ ,
\label{SM2}
\eea
i.e. the matter content of the Supersymmetric SM with
three generations of Higgses and vector-like colour triplets.


Let us remark that 
in these constructions
only the following patterns of matter with masses of order $M_{FI}$
are allowed:
%
\bea
a)\ n_3^{FI}=0\ ,\,\ \ n_2^{FI}=12 & \to & 3\times \left\{4(1,2)\right\}
\ ,
\nonumber
\\
b)\ n_3^{FI}=6\ ,\,\ \ n_2^{FI}=18 & \to & 
3\times \left\{(3,1)+(\bar 3,1)+6(1,2)\right\}
\ ,
\nonumber
\\
c)\ n_3^{FI}=12\ ,\ n_2^{FI}=24 & \to &
 3\times \left\{2[(3,1)+(\bar 3,1)]+8(1,2)\right\}
\ ,
\nonumber 
\\
d)\ n_3^{FI}=18\ ,\ n_2^{FI}=30 & \to & 
3\times \left\{3[(3,1)+(\bar 3,1)]+10(1,2)\right\}
\label{extra5}
\ .
\eea
Thus
for a given FI scale, $M_{FI}$, 
each one of the four patterns in eq.~(\ref{extra5})
will give rise to a different value
for $g_{GUT}$. Adjusting $M_{FI}$ appropriately,
we can always get $M_{GUT}\approx g_{GUT}\times 5.27\cdot 10^{17}$ GeV.
In particular this is so for 
$M_{FI}\approx 2\times 10^{16}$ GeV as shown in Fig.~1. 
It is remarkable that this number is within the allowed
range 
for the FI breaking scale as discussed above.
For the pattern in 
Fig.~1 corresponding to case {\it a)} we have 
$g_{GUT}\approx 1.1$, 
and therefore $M_{GUT}\approx 5.8\cdot 10^{17}$ GeV.


Of course, we cannot claim to have obtained the 
Heterotic String unification scale until we have shown
that the coupling $\alpha_1$ joins the other two couplings
at $M_{GUT}$.
The analysis becomes more involved now 
and a detailed account of this issue 
can be found in Ref.~\refcite{prediction}.
Let us just mention that the fact that the
normalization constant, $C$, of the $U(1)_Y$ hypercharge generator
is not fixed in these constructions 
as in the case of GUTs
(e.g., for
$SU(5)$, $C^2=3/5$) is crucial in order to obtain the unification
with the other couplings.

\vspace{0.5cm}

Summarizing,
the main characteristic of this scenario
is the presence at low energy of extra matter. In particular, we have
obtained that three generations of Higgses and vector-like colour
triplets are necessary.

Since more Higgs particles than in the MSSM are present,
there will be of course a much richer phenomenology. 
Note for instance that the presence of six Higgs doublets
implies the existence of sixteen physical Higgs bosons, where
eleven of them are neutral and five charged.  

Concerning the three generations of vector-like colour triplets, say 
$D$ and $\overline D$, 
they should acquire masses above the experimental limit $\cal{O}$(200
GeV).
This is possible, in principle,
through 
couplings with some of the extra singlets with vanishing
hypercharge, 
say $N_i$, 
which are usually 
left at low energies, even after the FI breaking.
For example, in the model of Ref.~\refcite{Casas2}, there are
13 of these singlets.
Thus couplings $N_iD\overline D$ might be present.
From the electroweak symmetry breaking, the fields $N_i$
a VEV might
develop. 
Note in this sense that the Giudice--Masiero 
mechanism to generate a $\mu$ term through the
K\"ahler potential is not available in prime orbifolds as $Z_3$.
Thus an interesting possibility to generate VEVs, 
given the large number of singlets present in orbifold models, 
is to consider couplings of the type $N_iH_uH_d$, similarly to the
Next-to-Minimal Supersymmetric Standard Model (NMSSM).
It is also worth noticing that some of these singlets might not have
the necessary couplings to develop
VEVs and then their fermionic partners might be candidates for right-handed 
neutrinos\footnote{Let us remark however, that right-handed neutrino 
superfields with R-parity breaking couplings 
of the type $N_iH_uH_d$ have been proposed
recently\cite{NewMSSM} to solve the $\mu$ problem.}.

For the models studied in Refs.~\refcite{Font,Giedt2} the extra colour
triplets have
non-standard fractional electric charge,
$\pm 1/15$ and $\pm 1/6$ respectively.
In fact, the existence of this kind of matter is a generic property
of the massless spectrum of supersymmetric models.
This means that they have necessarily 
colour-neutral fractionally charged states, since
the triplets bind with the ordinary quarks.
For example, the model with triplets with electric charge
$\pm 1/6$ will have mesons and baryons with charges 
$\pm 1/2$ and $\pm 3/2$.
On the other hand,
the model studied in Ref.~\refcite{Casas2}
has `standard' extra triplets, i.e. with electric charges
$\mp 1/3$ and $\pm 2/3$; these will therefore give rise
to colour-neutral integrally charged states.
For example, a $d$-like quark $D$ forms
states of the type $u\overline D$, $uuD$, etc.

Let us finally mention that a detailed discussion about the stability of these charged states,
how to solve possible conflicts with cosmological bounds, and
their production modes
can be found in Ref.~\refcite{prediction}.

\section{Quark and lepton masses and mixing angles}

Crucial ingredients in the above analysis were that all three
generations of
supersymmetric Higgses remain light 
($H^u_i$, $H^d_i$), $i=1,2,3$, and the FI breaking.
And, precisely, both ingredients favour to obtain the correct Yukawa
couplings at the renormalizable 
level\footnote{Let us recall that the major problem that one encounters
when trying to obtain models with entirely renormalizable Yukawas lies
at the phenomenological level, and is deeply related to obtaining the
correct quark mixing. Summarizing the analyses of Refs.~\refcite{test,Faustino}, for prime
orbifolds with the minimal Higgs content 
the space selection rules and the need for a fermion
hierarchy
forces the fermion mass matrices to be diagonal at the renormalizable
level.
Thus, in these cases the CKM parameters must arise at the
non-renormalizable level.
For analyses of non-prime orbifolds see Refs.~\refcite{test,Faustino,Ko1}.}. 
Namely,
having three families of Higgses introduces more Yukawa couplings,
and after the FI breaking some physical particles appear combined
with other states, and the Yukawa couplings are modified in a well
controlled way. This, of course introduces more flexibility in the
computation
of the mass matrices.

\begin{figure}[t]
\begin{center}
\epsfxsize=12pc 
\epsfbox{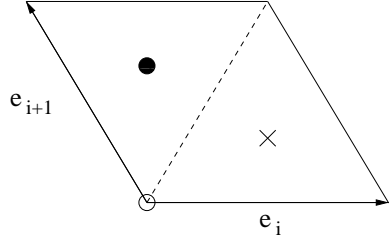}
\end{center} 
\caption{Two dimensional sublattices ($i=1,3,5$) 
of the $Z_3$ orbifold. The fixed point components are also shown.
\label{lattice}}
\end{figure}

Let us recall that the $Z_3$ orbifold
is constructed by dividing 
$R^6$ by the $[SU(3)]^3$ root lattice modded by the
point group ($P$) with generator $\theta$,
where the action of $\theta$ on the lattice basis is
$\theta e_i=e_{i+1}$, 
$\theta e_{i+1}=-(e_i+e_{i+1})$, with $i=1,3,5$.
The two-dimensional sublattices associated to $[SU(3)]^3$ are shown
in Fig.~\ref{lattice}.
In orbifold constructions, twisted strings appear 
attached to fixed points under the point group.
In the case of the $Z_3$ orbifold there are 27 fixed points under
$P$, and therefore there are 27 twisted sectors. 
We will denote the three fixed points of each two-dimensional
sublattice as shown in Fig.~\ref{lattice}. 
Thus the three generations arise 
because in addition to the overall factor of 3 coming from
the right-moving part of the untwisted matter, the twisted matter
come in 9 sets with 3 equivalent sectors on each one. 
Let us suppose that the two Wilson lines
correspond to the first and second sublattices.
The three generations correspond to move the third
sublattice component (x $\cdot$ o) of the fixed point keeping the
other two fixed.

As mentioned in the Introduction, we must arrange our up-and down-quark 
Yukawa couplings in order
to have specific off diagonal elements,
\begin{equation}
H_u \bar u_{L\alpha} \lambda_u^{\beta\gamma} u_{R\gamma}
+
H_d \bar d_{L\alpha} \lambda_d^{\beta\gamma} d_{R\gamma}\ .
\end{equation}
%
In principle this property arises naturally in the
$Z_3$ orbifold with two Wilson lines\cite{Ib.}$^-$\cite{Abel}.
For example, if the SU(2) doublet $H_u$ corresponds to (o o o), the three generations of 
(3,2) quarks to (o o (o, x, $\cdot$)) and the three generations
of ($\bar{3}$, 1) up-quarks to (o o (o, x, $\cdot$)),
then there are three couplings allowed from the space group
selection rule (the components of the three fixed points
in each sublattice must be either equal or different): 
$\lambda_{tt} H_u \bar t_{L} t_{R}$ associated to 
(o o o)(o o o)(o o o) with $\lambda_{tt}\sim$ 1,
$\lambda_{cu} H_u \bar c_{L} u_{R}$ associated to 
(o o o)(o o x)(o o $\cdot$) with
$\lambda_{cu}\sim e^{-T_5}$, and 
$\lambda_{uc} H_u \bar u_{L} c_{R}$
associated to (o o o)(o o $\cdot$)(o o x) with
$\lambda_{uc}\sim e^{-T_5}$.
In this simple example one gets one diagonal Yukawa coupling
without suppression factor and two off diagonal degenerate ones 
$\sim e^{-T_5}$, but other more realistic examples 
producing the observed structure of quark and lepton masses
and mixing angles
can be obtained
using 
three generations of Higgses\cite{Abel,teixeira2}.

Let us first study the situation 
before taking into account the effect of the FI breaking.
Consider for example the following assignments of observable matter to
fixed point components in the first two sublattices;
\bea
Q\ \,\,\,\, o\ o\ \,\,\,\,\,\,\,\,\,\,\,\, \ u^c\ \,\,\,\, o\ o
\ \,\,\,\,\,\,\,\,\,\,\,\, \ 
d^c\ \,\,\,\, x\ o
\nonumber
\\
H^u\ \,\,\,\, o\ o
\ \,\,\,\,\,\,\,\,\,\,\,\, \
H^d\ \,\,\,\, {\cdot}\ o
\ \,\,\,\,\,\,\,\,\,\,\,\, 
\label{assignments}
\eea
In this case the up- and down-quark mass matrices, assuming three 
different radii, are given by
%
\bea
M^u=gNA^u\ \,\,\,\,\ ,\ \,\,\,\, 
M^d=gN\varepsilon_1 A^d
\ ,
\label{quarkmasses}
\eea
where  
$g$ is the gauge coupling constant, $N$ is proportional
to the square root of volume of the unit cell for
the $Z_3$ lattice, and
\bea
A^{u}=\left( \begin{array}{ccc}
v_1^{u} & v_3^{u}\varepsilon_5  & v_2^{u}\varepsilon_5 \\
v_3^{u}\varepsilon_5  & v_2^{u} & v_1^{u}\varepsilon_5 \\
v_2^{u}\varepsilon_5  & v_1^{u}\varepsilon_5  & v_3^{u}
\end{array}\right)
\,\,\,\,\ , \,\,\,\,
A^{d}=\left( \begin{array}{ccc}
v_1^{d} & v_3^{d}\varepsilon_5  & v_2^{d}\varepsilon_5 \\
v_3^{d}\varepsilon_5  & v_2^{d} & v_1^{d}\varepsilon_5 \\
v_2^{d}\varepsilon_5  & v_1^{d}\varepsilon_5  & v_3^{d}
\end{array}\right)
\ .
\label{quarkmasses2}
\eea
Here
$v^u_i$, $v^d_i$ denote the VEVs of the Higgses 
$H^u_i$, $H^d_i$ respectively, and $\varepsilon_i=3\ e^{-\frac{2\pi}{3}T_i}$.
For example, for $T_5\sim 1.95$ one has
$\epsilon_5\sim 0.05$. 


The elements in the above matrices can be obtained straightforwardly.
For example, if the Higgs $H^u_1$ corresponds to 
(o,o,o), then since the three generations of (3,2) quarks $Q$ 
correspond to (o,o,(o,x,$\cdot$)) and the three generations
of ($\bar 3$,1) quarks $u^c$ to (o,o,(o,x,$\cdot$)),
there are only three allowed couplings,
\begin{eqnarray}
&&\mbox{(o,o,o)(o,o,o)(o,o,o)}\ ,\nonumber \\
&&\mbox{(o,o,o)(o,o,x)(o,o,$\cdot$)}\ , \nonumber \\
&&\mbox{(o,o,o)(o,o,$\cdot$)(o,o,x)}\ . \nonumber
\end{eqnarray}
\noindent The corresponding suppression factors are given by 1,
$\varepsilon_5$, $\varepsilon_5$ respectively,
and are associated with the elements 11, 23, 32 in the
matrix $M^u$. 

These matrices clearly improve the result obtained with only one Higgs
family. However, it is possible to show that 
although the observed quark mass ratios and Cabbibo angle can be
reproduced
correctly, the 13 and 23 elements of the CKM matrix cannot be obtained\cite{Abel}.
Fortunately, this is not the end of the story because the previous
result
is modified when one takes into account the FI breaking.
In particular, it will be possible to get the right spectrum and a CKM
with
the right form\cite{Abel,teixeira2}.

As discussed in the Introduction and Section~2, some scalars $C_i$
develop large VEVs
in order
to cancel the FI $D$-term generated by the anomalous $U(1)$. 
Thus many particles $\xi$ are expected to 
acquire a high mass because
of the generation of effective mass terms, and
in this way vector-like triplets and doublets and also singlets become 
heavy and disappear from the low-energy spectrum. 
This is the type of extra matter that typically 
appears in orbifold constructions. 
The remarkable point is that the SM matter remain massless,
surviving through certain combinations with other 
states\cite{Casas2,Font,Casas3}.
Let us consider the simplest example, a model with the
Yukawa couplings
\bea
C_1 \xi_1 f
\ ,
\,\,\
C_2 \xi_1 \xi_2 \, , 
\label{massive}
\eea
where $f$ denotes a SM field,
$\xi_{1,2}$ denote two extra matter fields (triplets, doublets or singlets),
and $C_{1,2}$ are the fields developing large VEVs
denoted by $\langle C_{1,2} \rangle = c_{1,2}$.
It is worth noting here that $f$ can be an $u^c$, $d^c$, $L$, 
$\nu^c$ or $e^c$
field, but not a $Q$ field. This is because in these orbifold models no 
extra (3,2) representations are present, and therefore the
Standard Model field $Q$ cannot mix with other representations through
Yukawas.

Clearly the `old' physical particle $f$ will combine with
$\xi_{1,2}$. It is now straightforward to diagonalise
the mass matrix arising from the mass terms in eq.~(\ref{massive}) to
find two very massive and one massless combination. 
The latter is given by
\bea
f'\equiv \frac{1}{\sqrt{|c_{1}|^{2}+|c_{2}|^{2}}}
\left(c_2^* f - c_1^* \xi_2\right)
\ .
\label{massless}
\eea
Notice for example that the mass terms (\ref{massive})
can be rewritten as
$\sqrt{|c_{1}|^{2}+|c_{2}|^{2}}\ \xi_1\xi'_2$,
where $\xi'_2\equiv
\frac{1}{\sqrt{|c_{1}|^{2}+|c_{2}|^{2}}}
\left(c_1 f + c_2 \xi_2\right)$.
Indeed the unitary combination is the massless field in eq.~(\ref{massless}).
The Yukawa couplings and hence mass matrices of the effective low energy 
theory are modified accordingly. 
For example, consider a model where we begin with a 
Yukawa coupling 
$H Q f$.
Since we have
\bea
f=\frac{1}{\sqrt{|c_{1}|^{2}+|c_{2}|^{2}}}
\left(c_2 f' + c_1^* \xi'_2\right)
\ ,
\label{oldfield}
\eea
then the `new' coupling (involving the light state) will 
be\footnote{We should add that the coupling
$H Q \xi_2$, which would induce another contribution to 
$H Q f'$, is not in fact allowed. For this to be the case
the fields $\xi_2$ and $f$ would have had 
to have exactly the same $U(1)^n$ charges.
This is not possible since different particles all have different
gauge quantum numbers.}
\[\frac{c_2}{\sqrt{|c_{1}|^{2}+|c_{2}|^{2}}} H Q f' \, .\]

The situation in realistic models is more involved since
the fields appear in three copies. 
All these effects modify 
the mass matrices of the low-energy effective theory
(see Eq.~(\ref{quarkmasses})), which, for the example studied
in Ref.~\refcite{Abel}, are now given 
by
%
\bea
{\cal M}^u=gNa^{u^c} A^u  B^{u^c}
\
\,\, ,\ \,\, 
{\cal M}^d=gN\varepsilon_1a^{d^c} A^d  B^{d^c}
\
\ ,
\label{finalud}
\eea
where 
\bea
A\  B
=
\left( \begin{array}{ccc}
v_1 \varepsilon_5 \beta\ & v_3\varepsilon_5  & v_2\alpha \\
v_3\varepsilon_5^2 \beta  & v_2 & v_1\alpha \\
v_2\varepsilon_5^2 \beta  & v_1\varepsilon_5\  & 
v_3\alpha/\varepsilon_5
\end{array}\right)
\ ,
\label{define}
\eea
and the parameters $a^{f}$, $\alpha$, and $\beta$ depend on\footnote{Note that
the $c_i$ are in general complex VEVs, 
and therefore they can give rise to a contribution to the CP
phase. This mechanism to generate the CP phase
through the VEVs of the fields cancelling the FI $D$-term was used
first, in the context of non-renormalisable couplings, in
Ref.~\refcite{test}. For a recent analysis, see Ref.~\refcite{Giedt:2000es}.} $c_{1,2}$\, 
$\epsilon_{1,3}$, and their possible values are discussed in
Ref.~\refcite{Abel}.
As shown in Ref.~\refcite{teixeira2}, for natural values of those parameters
and the VEVs, one can find configurations
that obey the electroweak symmetry breaking conditions, and
can account for the correct quark masses and mixings.

In addition to the magnitudes of the CKM matrix elements we also
require a CP violating phase. 
Although it has been shown that 
observable CP violation cannot be obtained at the
renormalizable level in odd order orbifolds\cite{oleg,Giedt:2000es}
for a minimal Higgs sector,
the above matrices having in addition to
the `mixing' of states three families of Higgses 
avoid this problem\cite{oleg2}.
Thus one possibility here
(in addition to the one already mentioned in footnote {\it i})
is to assume that the VEVs of the moduli have an imaginary phase,
which can occur when the flat moduli directions are lifted by 
supersymmetry breaking and find their minimum where the phases are 
non-zero\cite{Bailin1,oleg2,oleg3}. Such a phase feeds directly into 
$\varepsilon_5 $. It is easy to check that this phase is 
physically observable, and leads to a non-zero $\delta $ phase for the 
CKM matrix which is of order one.  

Let us finally mention that the correct masses
for charged leptons can be obtained following a similar approach, as
discussed
in Ref.~[\refcite{Abel}].
For neutrinos this turns out to be not sufficient, but
a see-saw mechanism arising in a natural way in 
orbifolds might solve the problem [\refcite{preparation}].

\section{Phenomenological viability of orbifold models with
three Higgs families}

The most challenging implication of an extended
Higgs sector is perhaps the
occurrence of tree-level FCNCs
mediated by the exchange of neutral Higgs states.
Clearly,
having six Higgs doublets (and thus six quark Yukawa couplings)
the transformations diagonalising the fermion
mass matrices do not diagonalise the Yukawa interactions.
Since experimental data is in good agreement with the SM predictions,
where such an effect is not present,
the potentially large contributions arising from the tree-level
interactions must be suppressed in order to have a model which is
experimentally viable. In general, the most stringent limit on the
flavour-changing processes emerges from the small value of the
$K_L-K_S$ mass difference\cite{McWilliams:1980kj}.

\begin{figure}[t]
  \begin{center}
    \begin{tabular}{cc}
      \epsfig{file=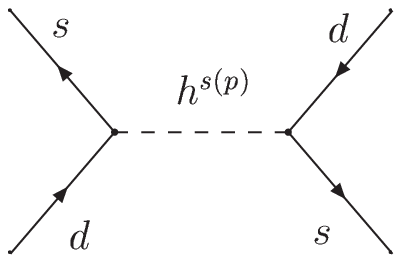,clip=}
      \hspace*{12mm}&\hspace*{12mm}
      \epsfig{file=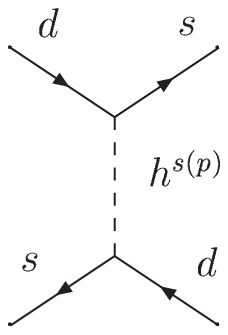,clip=}
      \\ & \\
      (a)\hspace*{12mm}&\hspace*{12mm} (b)
    \end{tabular}
    \caption{Feynman diagrams contributing to $\Delta m_K$ at
      tree-level. $h^{s(p)}$
denote scalar (pseudoscalar) Higgses.}
    \label{fig:effH}
  \end{center}
\end{figure}

A detailed discussion of FCNCs in multi-Higgs doublet
models was presented in Ref.~\refcite{Escudero:2005hk} (see also the references
therein). We summarise here 
some relevant points and apply the method to the
orbifold case\cite{teixeira2}, focusing on the neutral kaon sector and
investigating the tree-level contributions to $\Delta m_K$.
The latter is simply defined as the mass difference between the long-
and short-lived kaon masses,
\begin{equation}
\Delta m_K = m_{K_L} - m_{K_S} \simeq 2 \left| \mathcal{M}^K_{12}\right|=
2 \left| \langle \overline{K}^0 \left|
\mathcal{H}^{\Delta S=2}_{\text{eff}}\right| K^0 \rangle\right|\,,
\end{equation} 
where $\mathcal{H}^{\Delta S=2}_{\text{eff}}$ is the effective
Hamiltonian for the diagrams in Fig.~\ref{fig:effH}. Once all 
the contributions to $\mathcal{M}^K_{12}$ have been taken
into account, the prediction of this orbifold model
regarding $\Delta m_K$ should be compared with the experimental value, 
$(\Delta m_K)_{\text{exp}} \simeq 3.49 \times 10^{-12}$ MeV.

In Ref.~\refcite{teixeira2} 
the numerical approach was divided in two steps.
Firstly, one focus on the string sector of the model, and for each
point in the space generated by the free parameters of the orbifold
($\varepsilon_5, \alpha^f$), 
one derives the up- and down-quark mass matrices
and computes the CKM matrix.
Further imposing the conditions associated with electroweak
symmetry breaking, 
and fixing a 
value
for $\tan \beta$, one can
then determine the values of $g\,N$ and 
$\varepsilon_1$.
A secondary step requires specifying the several Higgs parameters,
which must obey the minimum criteria 
Finally, the last step comprehends the 
analysis of how each of the Yukawa patterns constrains
the Higgs parameters in order to have compatibility with the FCNC
data. In particular, we want to investigate how heavy the scalar and
pseudoscalar eigenstates are required to be in order 
to accommodate the observed
value of $\Delta m_K$.

Let us summarize the analysis of the orbifold parameter space by
commenting on the relative number of input parameters and number of
observables fitted. 
Working with the six Higgs VEVs ($v_i^{u,d}$), and the orbifold parameters 
$\varepsilon_1$, $\varepsilon_5$, $\alpha^{u^c}$ and $\alpha^{d^c}$,
one can obtain the correct electroweak symmetry breaking ($M_Z$), 
as well as the correct quark
masses and mixings (six masses and three mixing angles). 

In order to discuss now the tree-level FCNCs, let us remark that
the present orbifold model does not include a specific prediction 
regarding the Higgs sector. For instance, we have no
hint regarding the value of the several bilinear terms, nor
towards their origin. 
Concerning the soft breaking terms, the situation is similar. 
In the absence of further information, we merely assume that the 
structure of the soft
breaking terms is the 
usual one (see Ref.~\refcite{teixeira2} for further details), taking the Higgs soft
breaking masses and the $B \mu$-terms as free parameters (provided
that the electroweak symmetry breaking and minimisation conditions are verified).


In the absence of orbifold predictions for the Higgs sector
parameters, and 
motivated by an argument of simplicity, we begin our analysis by
considering textures for the soft parameters as simple
as possible. In particular, we arrive to four representative cases with
the following associated scalar and pseudoscalar Higgs spectra

\begin{itemize}
\item[(a)] 
$m^s = \{82.5, 190.6, 493.9, 515.9, 744.4, 760.2\}$ GeV\,;

$m^p = \{186.8, 493.9, 515.9, 744.4, 760.2\}$ GeV\,.
\item[(b)] 
$m^s = \{84.6, 213.9, 387.4, 560.8, 785.9, 879.1\}$ GeV\,;

$m^p = \{215.2, 387.3, 560.5, 785.9, 878.9\}$ GeV\,.
\item[(c)] 
$m^s = \{83.6, 292.9, 733.6, 785.9, 987.6, 1057.0\}$ GeV\,;

$m^p = \{291.1, 733.6, 785.9, 987.6, 1057.0\}$ GeV\,.
\item[(d)] 
$m^s = \{79.4, 121.5, 296.9, 354.3, 794.6, 808.8\}$ GeV\,;

$m^p = \{114.8, 296.9, 353.7, 794.6, 808.8\}$ GeV\,.
\end{itemize}

\begin{figure}[t]
  \begin{center} \hspace*{0mm}
      \psfig{file=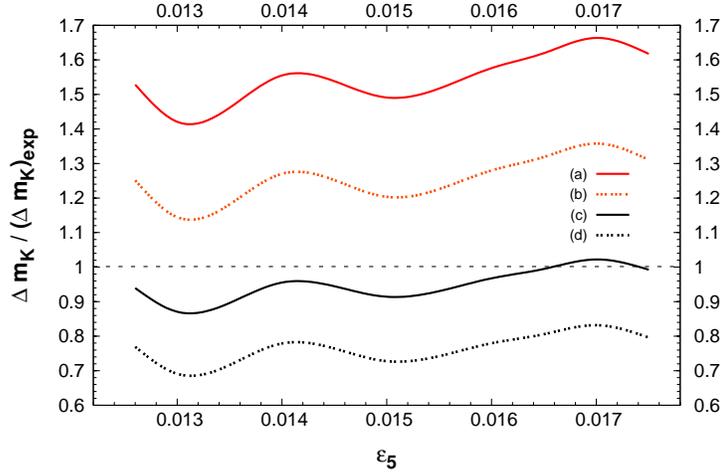,width=70mm,angle=270,clip=} 
    \caption{$\Delta m_K / (\Delta m_K)_{\text{exp}}$ as a function of 
$\varepsilon_5$ for $\tan \beta=5$. The Higgs parameters
      correspond to textures (a)-(d).}
    \label{fig:dmk}
  \end{center}
\end{figure}

In Fig.~\ref{fig:dmk} we plot the ratio 
$\Delta m_K/(\Delta m_K)_{\text{exp}}$ versus
$\varepsilon_5$, for cases (a)-(d), and 
$\tan \beta=5$.
All the points displayed comply with the bounds from the 
CKM matrix.
From Fig.~\ref{fig:dmk} it is clear that it is quite easy for the
orbifold model to accommodate the current experimental values for 
$\Delta m_K$. Even though the model presents the possibility of
important tree-level contributions to the kaon mass difference, 
all the textures considered give rise to contributions very close to the
experimental value. Although (a) and (b) are not in agreement with the
measured value of $\Delta m_K$, their contribution is within order of
magnitude of $(\Delta m_K)_{\text{exp}}$.
As seen from Fig.~\ref{fig:dmk}, with a
considerably light Higgs spectrum (i.e. $m_{h^0_i} < 1$ TeV), one is
safely below the experimental bound, as exhibited by cases (c) and (d). 
This is not entirely unexpected given the strongly hierarchical structure of
the Yukawa couplings (notice from Eq.~(\ref{define}) that $\lambda^d_{21}$
is suppressed by $\varepsilon_5^2$).

Let us finally mention that the analysis for other neutral
meson systems, $B_d$, $B_s$ and $D^0$,
can be carried out in an analogous way\cite{teixeira2}.

Additionally, and given the existence of flavour violating 
neutral Higgs couplings, and
the possibility of having complex Yukawa couplings, 
it is natural to have tree-level contributions to CP violation. In the
kaon sector, indirect CP violation is parameterised by
$\varepsilon_K$.
From experiment one has 
$\varepsilon_K = (2.284 \pm 0.014) \times 10^{-3}$.
A comparison of this quantity with the theoretical result 
in orbifold models can be found in Ref.~\refcite{teixeira2}.


\section{Conclusions}

We have attacked the problem of the unification of gauge couplings in
Heterotic String constructions. 
In particular, we have obtained that due to the Fayet-Iliopoulos scale, 
$\alpha_3$ and $\alpha_2$ cross at the right scale
when a certain type of extra matter is present. In this sense 
three families of supersymmetric Higgses and
vector-like colour triplets might be observed in forthcoming experiments.
The unification with 
$\alpha_1$ is obtained if the model has the
appropriate normalization factor of the hypercharge.
Let us recall that
although we have been working with explicit orbifold examples, our arguments
are quite general and can be used for other 
schemes where
the Standard Model gauge group with three generations of particles
is obtained,
since extra matter and
anomalous $U(1)$'s are generically present in string compactifications. 

Another advantage of 
these models is that they naturally predict three generations, and also 
that the three generations of Higgs fields give enough freedom
to allow an entirely geometric explanation of masses and mixings.
The Fayet-Iliopoulos mechanism plays also an important role here.
Namely, after the gauge breaking some physical particles appear
combined with other states, and the Yukawa couplings are modified in a
well
controlled way.

On the other hand, the presence of six Higgs doublets 
poses the potential problem of having tree-level FCNCs. 
By assuming simple textures for the Higgs free parameters, we have
verified for example that the experimental data on the neutral kaon mass
difference
can be
easily accommodated for a quite light Higgs spectra, namely 
$m_{h^0_i} \lesssim 1$~TeV. 


The presence of a fairly light Higgs spectrum, composed by a total of 21
physical states, may provide abundant experimental signatures at
future colliders, like the Tevatron or the LHC. In fact, flavour
violating decays of the form $h_i \to q \bar q$, or $h_i \to l^+ l^-$
may provide the first clear evidence of this class of models.

\section*{Acknowledgments}

This work was supported 
in part by the Spanish DGI of the 
MEC under Proyectos Nacionales FPA2006-05423 and FPA2006-01105;
by the Comunidad de Madrid under Proyecto HEPHACOS,
Ayudas de I+D S-0505/ESP-0346;
and also by the European Union under the RTN program  
MRTN-CT-2004-503369.






\end{document}